\documentstyle[aps,prl,twocolumn,epsfig]{revtex}

\title{Quantum computation with ions in thermal motion}

\author{Anders S\o rensen and Klaus M\o lmer\\
{\small
  Institute of Physics and Astronomy, University of Aarhus}\\
{\small DK-8000 \AA rhus C}}

\begin{document}
\draft
\maketitle
\begin{abstract}
We propose an implementation of  quantum logic
gates via virtual vibrational excitations
in an ion trap quantum computer.
Transition paths involving  unpopulated,
vibrational states interfere destructively 
to eliminate the dependence of rates and revolution frequencies on 
vibrational quantum numbers. As a consequence quantum computation becomes 
feasible with ions whos vibrations are strongly coupled to  a thermal
reservoir. 
\end{abstract}

Recently, methods to entangle states of two or several quantum systems 
in controlled ways have become subject of intense studies.
Such methods may find applications in
fundamental tests of quantum physics \cite{tests} and  in precision spectroscopy 
\cite{bollinger}; and they offer fundamentally new
possibilities in quantum communication and computing
\cite{issue}. A major obstacle to these efforts is decoherence of the 
relevant quantum states. In many proposed implementations of quantum computation  the
quantum bits (qubits) are stored in physical degrees of freedom with  long
coherence times, like nuclear spins, and decoherence is
primarily due to  the environment interacting with the channel
used to perform logic gates between qubits \cite{decoher}. In this Letter we present a
scheme which is insensitive to the interaction between the quantum channel
and the environment. Specifically, we  consider an implementation of
quantum computation in an ion trap, but we hope to stimulate similar ideas
to reduce decoherence in  other physical implementations. 

The ion trap was originally proposed by Cirac and Zoller  \cite{cirac} as a
system with good experimental (optical)  
access and control of 
the quantum degrees of freedom, and they  suggested an implementation
of the necessary ingredients in terms of one-bit and two-bit operations 
to carry out quantum computation. In the ion trap computer, 
qubits are represented by internal states of the ions. The number of qubits
equals the number of ions, and this system is  scalable to the
problem size in contrast to NMR quantum
computation which is only applicable with a limited number of qubits \cite{NMR}.

The ion trap method \cite{cirac} uses
collective spatial vibrations 
for communication between ions, 
and it requires that 
the system is restricted to the joint motional ground state of the ions. For
two ions this has recently been accomplished \cite{wineland}.
We present an alternative implementation of quantum gates that is
both insensitive to  the vibrational state   and robust against
changes in the vibrational motion occurring during operation, as long as the ions
are in  the Lamb-Dicke regime, {\it i.e.},
their spatial excursions are  restricted to a small fraction of the
wavelength of the exciting radiation. 
Our mechanism relies on
features of quantum mechanics that are often responsible for
``paradoxical effects": i) The vibrational degrees of freedom used
for communication in our scheme only enter virtually {\it i.e.},
although they are crucial as intermediate states in our  
processes, we never transfer population to states with different vibrational
excitation. ii) Transition paths involving different unpopulated,
vibrational states interfere destructively 
to eliminate the dependence of rates and revolution frequencies on 
vibrational quantum numbers.

\begin{figure}
\centerline{
\epsfig{file=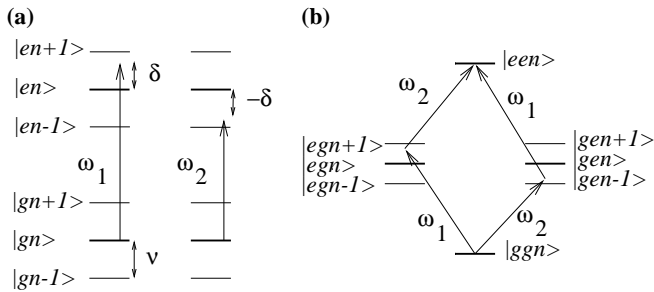,angle=270,width=\linewidth}}

\vspace{0.5cm}
\caption{Energy levels and laser detunings. {\bf (a)} Two ions with quantised
  vibrational motion are illuminated with lasers detuned close to the upper and lower sidebands.
  {\bf (b)} The ions oscillate in collective vibrational modes, and two
  interfering transition paths are identified.}
\label{detunings}
\end{figure}

Like in the original ion trap
scheme \cite{cirac}, we address each ion
 with a single laser, but quantum logic gates involving
two  ions are performed through off-resonant
laser pulses. For the laser addressing the first ion, we choose a detuning
close to the upper sideband, {\it i.e.},  close
to resonance with a joint vibrational and internal excitation of the ion. We choose the
detuning of the laser addressing the second ion  to be the
negative of the detuning of
the first laser,  see Fig. \ref{detunings} (a).
This laser setting couples the states  $|ggn\rangle \leftrightarrow
\{ |egn+1\rangle,|gen-1\rangle \} \leftrightarrow |een\rangle$, where the first
(second)  letter denotes the
internal state $e$ or $g$ of the first (second) ion and $n$ is the 
quantum number for the relevant vibrational mode of the trap. We choose 
the detuning from the sideband so large that the intermediate states
$|egn+1\rangle$ and $|gen-1\rangle$ are not populated in the
process. As we shall show below, the internal state transition is insensitive
to the vibrational quantum number $n$, and it may be applied even with
ions  which exchange vibrational energy with a surrounding reservoir.

If we tune the lasers sufficiently close to the sidebands we can neglect all
other vibrational modes and concentrate on  one  collective degree of vibrational
excitation of the ions \cite{strech}. In this case our system 
can be described  by the
following Hamiltonian 
\begin{eqnarray}
 H&=&H_0+H_{{\rm int}} \nonumber\\
 H_0&=&\hbar \nu (a^{\dagger} a+1/2)+\hbar \omega_{eg}\sum_i \sigma_{zi}/2 \nonumber\\
 H_{{\rm int}}&=&\sum_i \frac{\hbar\Omega_i}{2}
 (\sigma_{+i}e^{i(\eta_i(a+a^{\dagger})-\omega_i t)}+ h.c.),
 \label{hamilton}
\end{eqnarray}
where $\nu$ is the frequency and  $a^{\dagger}$ and $a$ are the ladder
operators of the vibrational  mode and  $\hbar w_{eg}$ is the
energy difference between the internal states $e$ and $g$.
Pauli matrices  $\sigma_i$ represent the internal degrees of freedom for the $i$'th
ion, and  $\omega_i$ and $\Omega_i$ are the frequency and
Rabi frequency of the laser addressing the $i$'th ion. [ In a practical
realization, one might use Raman transitions between low lying
states of the ions due to their long coherence time. By
appropriate redefinition of the symbols our
formalism also describes this implementation
\cite{raman}.]  We consider an ion trap
operating in the Lamb-Dicke limit, {\it i.e.} 
the ions are 
cooled to  a regime with vibrational quantum numbers $n$  ensuring that 
$\eta_i\sqrt{n+1}$ is well below unity (Note that this may still allow
n-values well above unity.) In our analytical calculations  we use an
expansion of  $H_{\rm int}$ to second order in $\eta_i$, but in our numerical
treatment we apply the exact Hamiltonian (\ref{hamilton}).

We wish to perform an operation on the mutual state of  two ions $1$ and $2$ 
selected freely within the  string of ions, and  we assume that
$\eta_1=\eta_2=\eta$ and $\Omega_1=\Omega_2=\Omega$. With the choice of
 detunings described above, the only energy conserving transitions are between
$|ggn\rangle$ and $|een\rangle$. The Rabi frequency $\tilde \Omega$ for the transition
between these states, via intermediate states $m$, can be  determined in
second order perturbation 
theory,
\begin{equation}
(\frac{\tilde \Omega}{2})^2 = \frac{1}{\hbar ^2} \bracevert \sum_m \frac{\langle
  een|H_{\rm int}|m\rangle \langle
  m|H_{\rm int}|ggn\rangle}{E_{ggn}+\hbar \omega_i -E_m}\bracevert ^2,
  \label{pertub}
\end{equation}
where the laser energy $\hbar \omega_i$ is the
energy of the laser addressing the ion which is excited in the intermediate
state $|m\rangle$.
If we restrict the sum to $|egn+1\rangle$ and 
$|gen-1\rangle$, we get
\begin{equation}
\tilde \Omega =- \frac{(\Omega \eta)^2}{2 (\nu - \delta)},
\label{omega}
\end{equation}
where $\delta=\omega_1-\omega_{eg}$ is the detuning of the laser addressing
the first ion \cite{deviation}. 

The  remarkable feature in Eq. (\ref{omega}) is that it contains no
dependence  on the vibrational quantum
number $n$. This is due to interference between the two paths indicated in
Fig. \ref{detunings}(b). If we take the path where ion No. 1 is excited first,
we have a factor of $n+1$ appearing in the enumerator ($\sqrt{n+1}$ from
raising and  $\sqrt{n+1}$ from lowering the vibrational quantum
number). In the other
path we obtain
 a factor of $n$. Due to the opposite detunings, the
denominators in Eq. (\ref{pertub}) have opposite signs and the $n$
dependence disappears when the two terms are subtracted. The coherent
evolution of the internal atomic
 state is thus insensitive to the vibrational quantum
numbers, and it may be observed with ions in any 
superposition or  mixture of 
vibrational states.

\begin{figure}
\centerline{
\epsfig{file=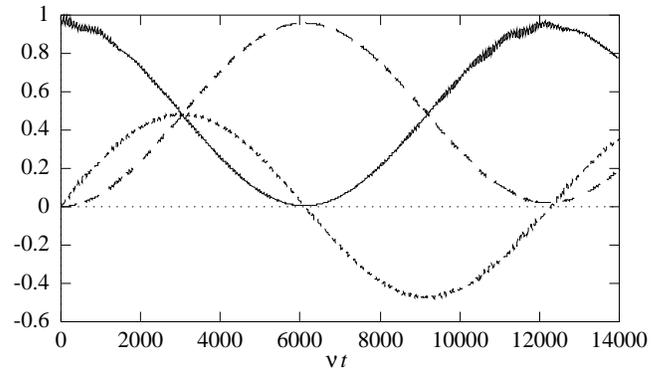,angle=270,width=\linewidth}}

 \caption{Rabi oscillations between $|gg\rangle$ and $|ee\rangle$. The
   figure shows the time evolution of the internal atomic density
   matrix elements  $\rho_{gg,gg}$ (full line), $\rho_{ee,ee}$ (long
   dashed line) and
   $Im(\rho_{gg,ee})$ (short dashed line). The magnitude of
   $Re(\rho_{gg,ee})$ is below 0.03 and is not shown. In the
   initial state, the  ions are in the internal ground
   state and a coherent vibrational state with mean excitation $\bar{n}= 2$.
   Parameters
   are $\delta=0.90~\nu$, $\Omega=0.10~\nu$ and $\eta=0.10$}
   \label{rabi}
\end{figure}

From the above arguments we expect to see perfect sinusoidal oscillations
between the population of the internal states $|gg\rangle$ and $|ee\rangle$.
 To confirm the validity of our
pertubative analysis we have performed a direct numerical integration of the
Scr{\"o}dinger equation with the Hamiltonian
(\ref{hamilton}) to all orders in $\eta$. We have considered a
situation, where both ions are initially in the internal ground state. For the
vibrational state, we have investigated a number of different states, including Fock,
coherent and thermal states, all yielding qualitatively similar results. The outcome of the
computation for a coherent state of vibrational motion can be seen in
Fig. \ref{rabi},
where we
show the 
evolution of relevant terms of the atomic internal state  density matrix $\rho_{ij,kl}=Tr_n
(\rho |kl\rangle \langle ij|)$, where $i,j,k,l=e$ or
$g$, and where $Tr_n$ denotes the partial trace over the unobserved
vibrational degrees of freedom. The figure clearly shows that we have Rabi oscillations between the
atomic states $|gg\rangle$ and $|ee\rangle$, and the values of the off diagonal element
$\rho_{gg,ee}$ 
 confirm that we have a coherent
evolution of the internal atomic state which is not  entangled with the
vibrational motion. 
Superimposed on the sinusoidal curves are small
oscillations with a high frequency due to off resonant couplings
of the type $|ggn\rangle \rightarrow |egn+1\rangle$,$|ggn\rangle
\rightarrow |gen-1\rangle$ and   $|ggn\rangle
\rightarrow |egn\rangle$. The magnitude of these oscillation and the
deviation from ideal transfer between $|gg\rangle$ and $|ee\rangle$  can be
suppressed by decreasing $\Omega$. 
 
The analysis given so far is sufficient for creation of internal state
entanglement, completely decoupled from the external motion of the ions. By
optical pumping we can prepare a state $\rho=|gg\rangle\langle gg|
\otimes \rho_{vib}$, and if
we apply radiation fields corresponding to a pulse of duration $T=\frac{\pi}{2|\tilde{\Omega}|}$, our
system is described by the density operator $\rho=|\psi\rangle\langle\psi|
\otimes \rho_{vib}$, where $|\psi\rangle$ is a maximally entangled EPR-state $\frac{1}{\sqrt{2}}
(|gg\rangle-i|ee\rangle)$.

\begin{figure}
\noindent
\centerline{
\epsfig{file=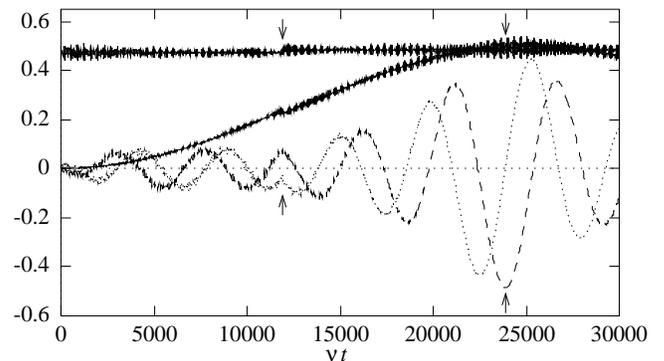,angle=270, width=\linewidth}}
 \caption{
Revival of coherence at $T_{inv}$. Initially ion 1 is in a
   superposition $\frac{1}{\sqrt{2}}(|g\rangle+|e\rangle)$ and ion 2 is in
   $|g\rangle$. 
The sign of $\delta$ is changed at $T_{inv}/2$
(left arrows), to ensure the perfect transition at $T_{inv}$
(right arrows). The full lines represent populations of $|ee\rangle$ and
$|eg\rangle$. The dotted and the dashed curves represent the real and
imaginary part of the coherence $\rho_{ee,eg}$. Parameters are
$\delta=0.90~\nu$,  $\Omega=0.05~\nu$ and $\eta=0.1$ }
 \label{echo}
\end{figure}

Since the states $|eg\rangle$ and $|ge\rangle$ 
do not fulfill any resonance condition one might expect that they are
unaffected by the laser pulses.  Due to  $n$-dependent
perturbations of  the energy levels by
the lasers this is, however, not the case. Keeping only the most important
terms, we get the energy shifts 
\begin{eqnarray}
 \Delta E_{ggn}&=&\Delta E_{een}=-\hbar \frac{(\eta\Omega)^2}{4}
 \frac{1}{\nu-\delta}\nonumber\\
 \Delta E_{egn}&=& \hbar \left(\frac{(\eta\Omega)^2}{2} 
 \frac{n}{\nu-\delta}-\frac{\Omega^2}{2\delta}\right)\nonumber\\
 \Delta E_{gen}&=& \hbar \left(- \frac{(\eta\Omega)^2}{2}
 \frac{n+1}{\nu-\delta}+\frac{\Omega^2}{2\delta}\right).
 \label{stark}
\end{eqnarray}
The energy shifts of the $|een\rangle$ and $|ggn\rangle$ are identical and
independent of $n$, but since the energy shifts of $|egn\rangle$ and $|gen\rangle$, depend on the
vibrational quantum number, the  time
evolution   introduces  phase factors $e^{-i\Delta E_{egn}t/\hbar}$,
which depend on $n$, and {\it e.g.}, at the time 
$t=T_{inv}=\frac{2\pi(\nu-\delta)} {\eta^2\Omega^2}$
where  $|gg\rangle$ end $|ee\rangle$ are inverted, factors
of $(-1)^n$ will tend to extinguish the coherence between internal states
$|ee\rangle$ and $|eg\rangle$. This coherence can be restored by a trick resembling photon
echoes \cite{photonecho}. Notice that the $n$ dependent part of $\Delta E_{egn}$ is minus the 
$n$ dependent part of $\Delta E_{gen}$. If at any time $T/2$ we change the sign
of the laser detuning $\delta$,  
phase components proportional to $n$ will begin to
rotate in the opposite direction and at time $T$ we will have a
revival of the coherence. 
This is confirmed by our numerical solution of the Schr{\"o}dinger equation
presented in Fig. \ref{echo}, where we change
the laser detunings at
the time $T_{inv}/2$ and at the time $T_{inv}$ we have  completed the
transfer $\frac{1}{\sqrt{2}}(|gg\rangle+|eg\rangle) \rightarrow
\frac{1}{\sqrt{2}}(-i |ee\rangle+|eg\rangle)$.

No particularly
demanding assumptions have been made for the experimental parameters.
With a vibrational frequency $\nu/2\pi = 200$ kHz, 
the transition shown in Fig. \ref{rabi}, require Rabi frequencies $\Omega/2\pi$
of modest  $20$ kHz, and the evolution from $|gg\rangle$ to $|ee\rangle$ is
accomplished in 5 ms. To be relevant for real computational tasks it is
necessary that our evolution is robust against decoherence
effects on this long time scale. An important source of decoherence is
heating of the vibrational motion, and it is a major asset
of our proposal that it can be made 
insensitive to the interaction with the environment: The arguments leading
to Eq. (\ref{omega}) do not 
require that the ions remain in the same vibrational state, and the
coherent oscillation from $|gg\rangle$ to $|ee\rangle$ may still be observed when
the vibrational motion  exchange energy with a thermal reservoir. The
photon echo trick, however, is sensitive to heating: If the vibrational
quantum number $n$ change  its value at
the time $T/2$ where the detunings are inverted, 
 the second half of the gate, will no longer revert the
phase evolution due to the new value of $n$ and coherence is lost.  If
instead the detunings are  inverted
$N$ times during a gate, the erroneous phase
will only be on the order of the phase evolution  in time $T/N$, and the
effect of the heating is reduced.

Rather than inverting the detunings a large number of times, we suggest
to continuously apply lasers with
both detunings $\pm \delta$ on both ions. With two fields of opposite
detunings and identical Rabi  frequency $\Omega$
there are two contributing paths in addition to the two paths in
Fig. \ref{detunings}. The contribution from the two additional paths are
identical to the two original paths and the only modifications to the
$|gg\rangle \leftrightarrow |ee\rangle$ Rabi frequency in Eq. (\ref{omega}) is
multiplication by a factor of two. With bichromatic fields there also
exists a  resonant transition from $|eg\rangle$ to $|ge\rangle$. The Rabi
frequency of  this transition is the negative of the Rabi frequency from
$|gg\rangle$ to $|ee\rangle$ and the evolution will be described by
\begin{eqnarray}
 |gg\rangle &\rightarrow& \cos(\frac{\tilde{\Omega}T}{2})|gg\rangle +i~
 \sin(\frac{\tilde{\Omega}T}{2})|ee\rangle \nonumber\\
 |ee\rangle &\rightarrow& \cos(\frac{\tilde{\Omega}T}{2})|ee\rangle +i~
 \sin(\frac{\tilde{\Omega}T}{2})|gg\rangle \nonumber\\
 |ge\rangle &\rightarrow& \cos(\frac{\tilde{\Omega}T}{2}) |ge\rangle - i~
 \sin(\frac{\tilde{\Omega}T}{2}) |eg\rangle\nonumber\\
 |eg\rangle &\rightarrow& \cos(\frac{\tilde{\Omega}T}{2}) |eg\rangle -i~
 \sin(\frac{\tilde{\Omega}T}{2})|ge\rangle. 
 \label{bievolution}
\end{eqnarray}

\begin{figure}[htbp]
  \begin{center}
   \begin{minipage}{4.55cm}
    \epsfig{file=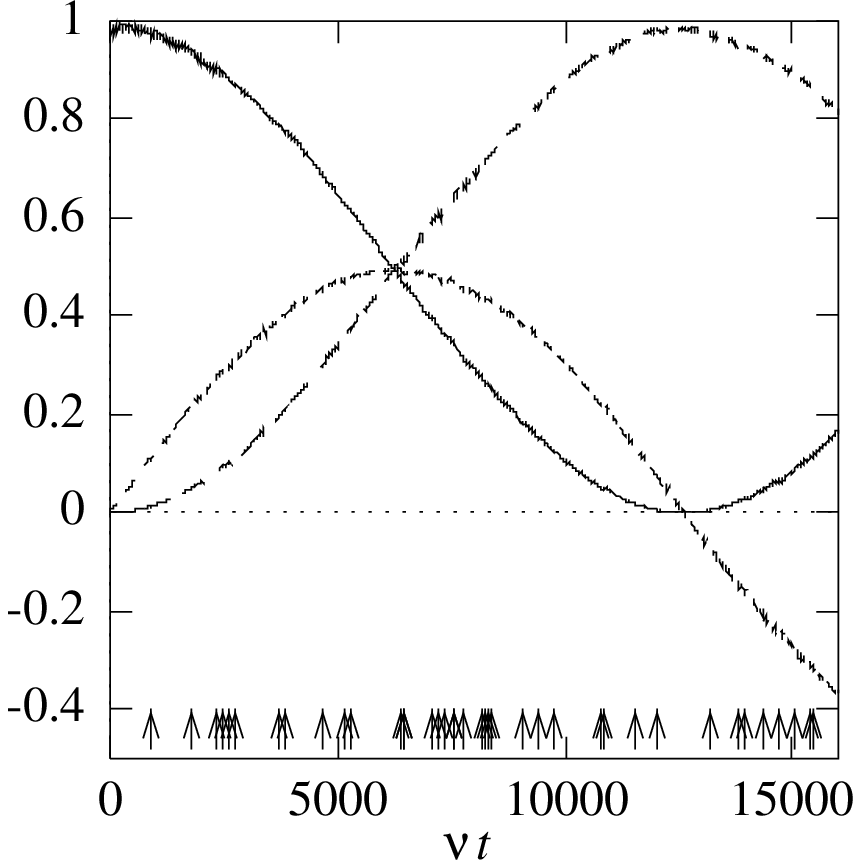,width=4.45cm,angle=270}
   \end{minipage}
   \begin{minipage}{3.95cm}
    \vspace{0.1cm}
    \epsfig{file=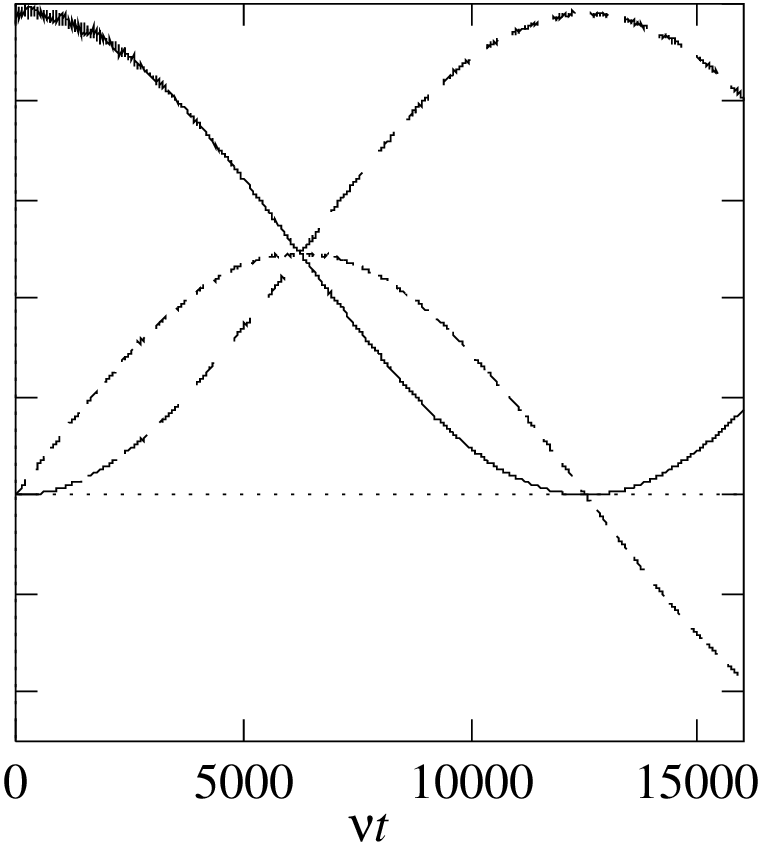,width=4.35 cm,angle=270}
   \end{minipage}
  \end{center}
  \caption{Rabi oscillations in a heating trap. The left panel shows the
    result of a single Monte Carlo realization with a total of 39 jumps
    occuring at times indicated by the
    arrows. The right panel is an average   over 10 realizations. The curves
    represent $\rho_{gg,gg}$ (full line), $\rho_{ee,ee}$ (long dashed) and
    $Im(\rho_{gg,ee})$ (short dashed). Parameters
    are $\delta=0.90~\nu$, $\Omega=0.10 ~\nu$, $\eta=0.1$,
    $\Gamma=2*10^{-4}~\nu$ and $n_{\rm term}=2$.}
  \label{heat}
\end{figure}

To validate that the evolution in Eq. (\ref{bievolution}) is in fact stable
against heating we introduce a  thermal
reservoir described by relaxation operators
$c_1=\sqrt{\Gamma(1+n_{{\rm therm}})}a$ and
$c_2=\sqrt{\Gamma n_{\rm therm}}a^\dagger$, where  $\Gamma$ characterizes the
strength of the interaction and $n_{\rm therm}$ is the mean vibrational number
in thermal equilibrium. We analyse the dynamics of the system using Monte
Carlo wavefunctions \cite{montecarlo}, which evolve with a
non Hermitian Hamiltonian interrupted by jumps at random times. 
 The result of the computation can be seen in
Fig. \ref{heat}, where we show (left) the result of a single Monte Carlo
realization with quantum jumps indicated by arrows and (right) the average over
10 realizations. In the figure we have chosen 
$n_{\rm therm}=2$. This rather low value could represent a heating
mechanism counteracted by laser cooling on a particular ion reserved for
this purpose. In the simulations  34  vibrational quanta
are exchanged with the reservoir on average, and we wish to emphasize
that with the proposed scheme the gate is almost unaffected even
though the duration of the gate is much longer than the coherence time of
the channel used to communicate between the qubits.

It has been proven that any unitary
evolution involving a number of qubits can be constructed using only single
qubit operations and a simple universal quantum gate like for instance the
two-qubit control-not operation \cite{gates}. With the evolution described
by Eq. (\ref{bievolution}) a control-not operation is
created by the following sequence of operations: $P_1$, ${P_2}^{-1}$, $H_2$, $R$,
$P_1$, $H_1$, $P_1$, $R$ 
and $P_2$, where $R$ is the evolution in Eq. (\ref{bievolution}) with
$T=\frac{\pi}{2|\tilde{\Omega}|}$, $P_i$ is a $\pi/2$ phase change of
$|e\rangle$ in ion $i$, and $H_i$ is a Hadamard transformation on ion
$i$. With ready access to one-qubit
operations in the ion trap we thus have available the ingredients for
successful quantum computation.



We note that with only two ions in the trap, the use of bichromatic light leads to
the evolution (\ref{bievolution}) even without individual optical access,
and if many ions  are illuminated by the same field, multi particle
entanglement is created \cite{cat}. As a further remarkable feature of our implementation, we note how
easy it is to simultaneously operate on different pairs of ions:
If we wish to apply our gate on the pairs $(i,j)$ and $(k,l)$,
we simply illuminate ions $i$ and $j$ with fields of detunings
$\pm \delta_{ij}$ and ions $k$ and $l$ with another 
pair of detunings $\pm \delta_{kl}$. The only resonant transitions
are the ones of the two desired gates and, however mind boggling this may
be, the virtually excited but never populated vibrational mode has been
used for two processes at the same time.

{\it Note added:} Since submission of this work, two proposals for
computing with vibrationally excited ions have appeared. \cite{poyatos}
uses widely separated vibrational states, and can hardly be generalized
beyond  2 or 3 qubits; \cite{milburn} involves dynamics which is
conditioned on the vibrational state and is not applicable during heating.


\begin{thebibliography}{99}
\bibitem{strech} Numerical calculations retaining two collective vibrations
  for a two ion system have confirmed the validity of the omission of the far off-resonant
  vibration modes.
\bibitem{deviation} If the additional coupling to the lower (higher) sideband of
  ion 1 (2) is included  $2(\nu-\delta)$ in Eq. (\ref{omega}) is
  replaced by $(\nu^2-\delta^2)/\nu$.
\end{thebibliography}
\end{document}